# ReGraphX: NoC-enabled 3D Heterogeneous ReRAM Architecture for Training Graph Neural Networks

Aqeeb Iqbal Arka*, Biresh Kumar Joardar†, Janardhan Rao Doppa*, Partha Pratim Pande*, and Krishnendu Chakrabarty†

*School of EECS, Washington State University
Pullman, WA 99164, USA.
{aqeebiqbal.arka, jana.doppa, pande}@wsu.edu

†Department of ECE, Duke University
Durham, NC 27708, USA.
{bireshkumar.joardar, krish}@duke.edu

*Abstract*— Graph Neural Network (GNN) is a variant of Deep Neural Networks (DNNs) operating on graphs. However, GNNs are more complex compared to traditional DNNs as they simultaneously exhibit features of both DNN and graph applications. As a result, architectures specifically optimized for either DNNs or graph applications are not suited for GNN training. In this work, we propose a 3D heterogeneous manycore architecture for on-chip GNN training to address this problem. The proposed architecture, ReGraphX, involves heterogeneous ReRAM crossbars to fulfill the disparate requirements of both DNN and graph computations simultaneously. The ReRAM-based architecture is complemented with a multicast-enabled 3D NoC to improve the overall achievable performance. We demonstrate that ReGraphX outperforms conventional GPUs by up to 3.5X (on an average 3X) in terms of execution time, while reducing energy consumption by as much as 11X.

## I. INTRODUCTION

Predictive analytics over graph data enables diverse real-world applications in various domains, including biology, social networks, and recommendation systems [1]. A key challenge here is to learn good representations (i.e., discriminative features) over nodes, edges, and graphs. Recent advances in graph neural networks (GNNs) have addressed this challenge with considerable success [2]. Unlike deep neural networks (DNNs) over regular structures such as images (e.g., CNNs) and sequences (e.g., RNNs), the overall computational process in GNNs is extremely challenging due to the relational structure of graphs. GNNs perform an iterative neighborhood aggregation operation, where each node aggregates features of its neighbors to compute the new features [3]. After $k$ iterations, the transformed feature vector of a node captures the relational structure information within this node's $k$-hop neighborhood. The representation of an entire graph is computed by pooling the feature vectors of all the nodes. More specifically, GNNs combine two distinct types of operations: 1) Vertex-centric computations, which involve trainable weights and are similar to conventional DNNs, and 2) Edge-centric computations, which involve accumulating neighbor information along the edges of the graph [3] [4]. Hence, the GNN training process exhibits features of both DNN training (compute-intensive) and graph computation (heavy data exchange). Neither conventional CPU- nor GPU-based systems are tailored for this family of emerging applications. Therefore, efficient hardware platforms to enable GNN training/inference are necessary.

The vertex- and edge-centric computations in GNNs can be decomposed into a set of multiply-and-accumulate (MAC) operations. It is well known that resistive random-access memory (ReRAM) can implement MAC operations very efficiently [5]. In addition, ReRAMs compute in-memory; this reduces the amount of communication (data transfers) between computing cores and the main memory for GNN training/inference. Hence, ReRAM-based architectures are naturally suited for accelerating GNN training/inference (compared to GPUs). However, existing ReRAM-based architectures are optimized specifically for either DNNs (e.g. [6], [7]) or graph computations (e.g. [8]); they are not effective for GNN training as they exhibit characteristics of both DNNs and graph computations. Hence, there is a pressing need for a heterogeneous architecture that synergistically combines the design principles from both DNNs and graph analytics.

Unlike traditional DNNs, GNNs involve heavy data exchange due to message-passing operations to accumulate neighbor information in a recursive approach [9]. This gives rise to significant data exchange among the ReRAM tiles, which limits the overall achievable performance. We show in this work that data exchange during GNN training predominantly exhibits *many-to-one-to-many,* and multicast characteristics. Traditional planar (two-dimensional) architectures are not suitable for GNN training due to the large physical separation between the tiles. This will result in significant amount of long-range communication causing a performance bottleneck [10]. To reduce the performance bottlenecks in planar designs, three-dimensional network-on-chip (3D NoC) based architectures should be used. By stacking planar dies on top of one another and connecting them with vertical links, the communication latency can be greatly reduced. Prior work has shown that 3D NoCs enable the design of a low latency and high throughput communication backbone for manycore chips [11]. In addition, 3D NoC enables high-throughput multicast support [12]. As a result, 3D NoC is more suited for long-range, multicast, and many-to-one-to-many traffic patterns [12], [13]. Hence, we conjecture that 3D NoC is a suitable communication backbone to enable the proposed ReRAM-based architecture for training GNNs.

In this paper, we present the design of a 3D NoC-enabled manycore architecture, referred as *ReGraphX*, for training GNNs. The proposed manycore architecture consists of (a) heterogeneous ReRAMs as the Processing Elements (PEs) to accelerate the large number of MAC operations in GNNs and (b) a high-throughput NoC architecture as the communication backbone. The main contributions of this work include:

- We undertake an in-depth study of the computation and communication patterns when GNNs are trained on ReRAM-based platforms. This study motivates and influences the design of an efficient NoC architecture.
- We design an energy efficient and high-performance 3D NoC architecture by taking into consideration the on-chip traffic pattern associated with GNN training. The 3D NoC architecture enables high performance and energy efficient GNN training.

To the best of our knowledge, this is the first work that proposes a heterogeneous ReRAM-based manycore architecture enabled by 3D NoC for training GNNs. The rest

This work was supported, in part by the US National Science Foundation (NSF) grants CNS-1955353, CNS-1564014, CNS-1955196 and USA Army Research Office grant W911NF-17-1-0485. This material is also based upon work supported by the National Science Foundation under Grant # 2030859 to the Computing Research Association for the CIFellows Project.

of the paper is organized as follows. Section II describes relevant prior work. In Section III, we discuss the salient features of GNNs, especially the traffic patterns that must be considered for NoC design. In Section IV, we introduce the proposed ReGraphX architecture, the role of the 3D NoC, and the GNN layer mapping strategy. Section V presents experimental results and analysis. We conclude the paper in Section VI.

## II. RELATED PRIOR WORK

### A. ReRAM-based architectures

ReRAMs can be used as memory and also to perform *in-situ* MAC operations [5]. Both DNN and graph applications involve significant amount of MAC operations. Hence, ReRAM-based accelerators for both DNN training and inference have been extensively studied [6] [7]. Moreover, ReRAM based graph accelerators have been shown to significantly outperform CPU- or GPU-based systems both in terms of execution time and energy [8] [14] [15]. However, these solutions focus mainly on accelerating the computation. As mentioned earlier, GNNs involve heavy communication, which limits the maximum achievable performance. In addition, all these ReRAM-based accelerators are fine-tuned either for DNN training/inference or graph applications. Our work addresses a key shortcoming of the state-of-the art by proposing an architecture that caters to GNN training.

### B. Hardware for GNN Computation

GNNs exhibit a unique computational kernel that is a combination of traditional DNNs at vertices and graph analytics at edges [4]. However, GNN computation is memory intensive for large-scale graphs, which necessitates use of efficient graph partitioning [16]. This divide-and-conquer approach enables scalable GNN training over large graphs with high accuracy and speed [16]. The design of hardware architectures using commodity processors, FPGAs and custom ASICs has been considered in recent work [2] [9] [10]. However, all these architectures primarily focus on GNN inference but *not* training. The training of GNNs is considerably more challenging due to additional data exchange between the forward and backward phases. Moreover, all these architectures are limited to relatively small graph structures, which do not require large amounts of memory and computation. In contrast, this paper is focused on a single-chip architecture enabled by 3D integration that leverages the benefits introduced by ReRAM-based processing elements (PEs) to design a GNN training accelerator. Moreover, the proposed architecture leverages graph partitioning to enable acceleration of training GNNs with large-scale graphs that contain millions of nodes.

## III. COMPUTATIONAL KERNEL FOR TRAINING GNNS

In this section, we present the salient features of GNN training. Specifically, we study its computation and communication characteristics to motivate the proposed ReRAM-based architecture. Fig. 1 shows the different computational and communication characteristics of a GNN. As shown in Fig. 1(a), a graph consists of (i) vertices: each vertex $V$ can be represented using a feature vector (denoted as the vector $X_V$ in Fig. 1(a)); and (ii) edges, which are represented as an adjacency matrix ($Adj$). The feature vector $X_V$ characterizes a given node $V$ of a graph while $Adj$ indicates the vertex connectivity.

A GNN consists of multiple cascaded neural layers that in turn can be divided into two sub-layers (or sub-tasks): (i) Vertex-computation layer (V-layer; shown in Fig. 1(b)), and (ii) Edge-computation layer (E-layer; shown in Fig. 1(c)). The V-layer involves multiplying the feature vector of the nodes/vertices ($X_V$) with the weight matrices ($W$). It is like MAC operations in a traditional DNN as shown in Fig. 1(b). Hence, this operation can be efficiently implemented using ReRAM-based architectures [6] [7]. The output of the V-layer is the updated vertex feature vector ($Y_V = W \cdot X_V$). The E-layer involves the accumulation of the updated neighbor information once the V-layer is completed. This is done via the graph edges and is similar to the message-passing operation between all one-hop neighbors, as shown in Fig. 1(c) [4]. Interestingly, this operation can also be decomposed as a matrix-multiplication operation as shown in Fig. 1(c). The E-layer computation can be envisioned as a sparse matrix-vector multiplication (SpMV) involving $Adj$ (sparse matrix [2]) and the *updated* vertex feature vectors ($Y$) as shown in Fig. 1(c). For instance, node $A$ in Fig. 1 should accumulate the updated information from its 1-hop neighbors: node $B$ and node $C$ in the E-layer, which can be alternatively accomplished by multiplying the first row of $Adj$ and the vertex feature matrix as shown in Fig. 1(c). Hence, as Fig. 1 shows, both the V- and E- layers can be decomposed as MAC operations which can be implemented using ReRAMs [5].

Next, Fig. 1(d) shows an example GNN with three neural layers, i.e., three V-E layer pairs. As mentioned earlier, the V-layers have trainable weights similar to a standard DNN. Each V-layer has a unique set of weights which need to be mapped/assigned to different sets of PEs (V-PEs) for computation (discussed in more detail in the next section). The E-layer requires only the graph adjacency matrix, which is represented by $Adj$. The $Adj$ matrix is *fixed for a given graph* and is mapped to another set of PEs (E-PEs). Similar to V-PEs, we can have separate E-PEs for each neural layer of GNN. However, as $Adj$ is constant for a given graph, the different E-PEs will store the same information and is unnecessary. As a result, the E-PEs need to be shared by all the neural layers. As discussed earlier, the output of the V-layer is used as input for the next E-layer and so on. This results in a *many-to-one-to-many* communication pattern as multiple sets of V-PEs communicate with the same set of E-PEs as shown in Fig. 1(d). Without a suitable interconnection backbone, the many-to-one-to-many communication can overwhelm the training process, resulting in a performance bottleneck. Hence, a suitable NoC is needed for high-performance training of GNNs.

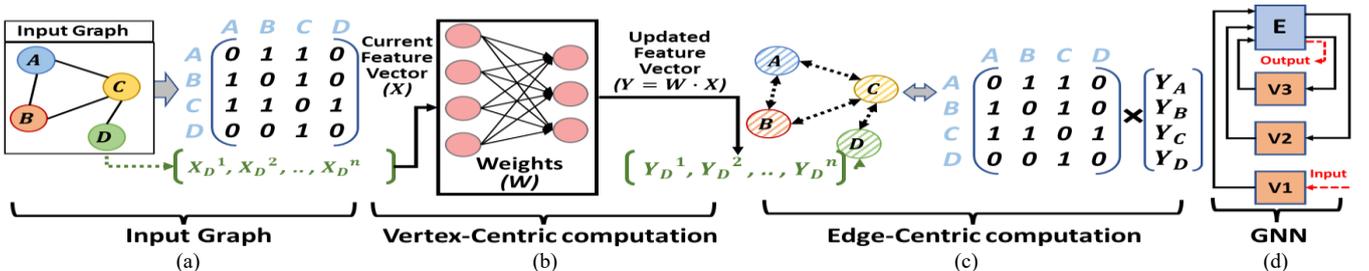

Fig. 1. Illustration of the computational components of a GNN: (a) Input graph represented as node features ($X_i$) and adjacency matrix ($Adj$), (b) Vertex-centric computation layer (V), (c) Edge-centric computation layer (E); Both V- and E-layers together, constitute a neural layer of GNN, and (d) Overall GNN structure with three neural layers as an example; The arrows indicate the data communication pattern in a GNN. Note that each V-layer has its unique set of weights (like traditional DNNs) which need to be mapped to different PEs. However, the E-layer depends on $Adj$ only which is fixed for a given input graph. This results in a many-to-one communication where all the PEs responsible for V-layer computations communicate with the same PEs storing the $Adj$ matrix.

## IV. OVERALL REGRAPHX ARCHITECTURE

In this section, we first present the key features of the ReGraphX architecture. Fig. 2 shows the proposed ReGraphX architecture.

### A. Role of Heterogeneity

To accelerate the V-layer (which resembles a conventional DNN), we utilize the ReRAM-based tiled architecture for DNNs [6]. The ReRAM tiles consist of 128x128 ReRAM crossbars (V-PEs) and associated peripheral circuits. The weights ($W$ in Fig. 1(b)) are mapped to the ReRAM cells for high-throughput MAC operations. We do not describe how weights are mapped to individual ReRAM cells in detail for the sake of brevity. However, the E-layers in GNNs require the storage of the $Adj$ matrix (Fig 1(c)), which is sparse. Typically, a significant portion (over 90%) of $Adj$ are zeros. Multiplication with zeros is redundant and should be avoided. This can be accomplished by dividing the larger $Adj$ matrix (of size $N \times N$; $N$ being the number of nodes in the graph) into smaller sub-matrices (of size $M \times M$, where $M << N$). Then, any sub-matrix with all $M \times M$ entries being zero is discarded while the rest are mapped to $M \times M$ ReRAM crossbars (similar to DNN weights). Note that this method does not eliminate all zero entries of $Adj$; rather it reduces the number of zeros stored leading to lower E-PE requirements and energy dissipation.

We show in Fig. 3 that larger ReRAM crossbars (as in [6]) are not suited for this task. Fig. 3 shows the number of zeros that are stored (after the reduction step discussed above) for two different ReRAM crossbar sizes ($M \times M$) for the three datasets *PPI*, *Reddit*, and *Amazon2M* considered in this work. As we can see from Fig. 3, larger ReRAMs store up to 7X more zeros than their smaller counterparts, which leads to a significantly higher number of redundant multiplications. This leads to high energy consumption and lower throughput as resources are wasted on redundant multiply-by-zero operations. In addition, more ReRAM cells are necessary to store the extra zeros (higher area/hardware overhead). Hence, smaller ReRAM crossbars are more efficient for the E-layer computations. Overall, ReGraphX is a heterogeneous architecture that consists of two types of PEs (ReRAMs in this case): larger ReRAM crossbars (128x128) for the V-layer computations (V-PEs) and smaller crossbars (8x8) for the E-layers (E-PEs) as shown in Fig. 2. It should be noted that even smaller ReRAM sizes can also be used for E-PEs [14]. In this work, we adopt (without any loss of generality) the 8x8 tile architecture for E-PE following recent trends [8] [15].

### B. Role of 3D NoC

As shown in Fig. 1(d), the V-PEs send the updated node features ($Y$ in Fig. 1) to the E-PEs for further processing. The amount of data communicated between the V-PEs and the E-PEs is proportional to the total number of nodes, which considering all the features in the input graph, is often very high. For example, each image in the popular ImageNet dataset (often used for evaluating DNNs) consists of 150K entries (image size: 224x224x3) which is an order of magnitude smaller than even the smallest input (sub-)graph considered in this work (1.6 million entries for PPI dataset; more details in Section V). This results in significant amount of data exchange during GNN training. In addition, the data communication exhibits many-to-one-to-many patterns, where multiple sets of V-PEs communicate with the same set of E-PEs, and vice versa. Moreover, training involves data sharing between the forward- and backward-phase computations of each layer; often the backward-phase computations are implemented on separate set of ReRAMs, as described in [7]. Overall, this results in the output of layer $L_i$ being sent to: (a) PEs responsible for the next layer $L_{i+1}$, and (b) the PEs responsible for the backward phase of layer $L_i$. Therefore, there is a significant amount of multicast traffic on top of the many-to-one-to-many traffic.

Traditional planar architectures are not suited for such traffic patterns. The heavy many-to-one-to-many and multicast traffic, as well as the large physical separation between PEs, imposes a significant amount of long-range communication requirement in planar architectures. As a result, the communication backbone quickly becomes a performance bottleneck. In addition, the multi-hop nature of a planar-mesh NoC leads to higher communication latencies [13], which is not desirable for training GNNs. A 3D architecture can alleviate this problem and be a powerful enabler for the ReGraphX architecture. By stacking multiple layers above each other, the physical distance between PE tiles is reduced significantly [13]. In addition, 3D NoCs enable high-performance multicast support [12]; both these characteristics are beneficial for GNN training. In this work, we use a 3D mesh NoC as interconnection backbone because it can support 3D tree multicast. We consider tree-multicast in this paper without any loss of generality. Other multicast-aware routing schemes can also be used for ReGraphX.

Overall, ReGraphX consists of three physical layers (tiers), as shown in Fig. 2. We need more E-PEs than V-PEs as the number of trainable weights in the V-Layer (few hundred thousands) is smaller than the number of entries in the $Adj$ matrix (which can run into millions) that need to be stored for the E-layer for an input graph. Hence, ReGraphX has two E-PE tiers and one V-PE tier. The middle tier consists of V-PEs and has one-hop connections facilitated by the vertical links in both the vertical directions. The top and bottom layers include the E-PEs that store the *Adj* matrix. *This sandwiched structure* (Fig. 2) is critical as it provides one-hop access between the V-PEs and the E-PEs in both directions (a.k.a. vertical traffic; shown in Fig. 1(d)). The intra V-PE communication (a.k.a. planar traffic) due to data sharing between the forward and backward phases is addressed by the multicast-aware mapping policy, which maps communicating neural layers to nearby PEs. The E-layer does not have any learnable/trainable parameter. Hence, there is no separate forward-backward traffic as in the case of V-PEs. However, there will be planar traffic in the E-layer as *Adj* is distributed across multiple E-PEs. Overall, the NoC and the mapping policy complement each other for high-performance GNN training. Note that, three tiers are used as an example only, to show the efficacy of 3D architectures for GNN training. More tiers can also be used. However, adding more tiers can lead to thermal issues and investigating thermal-aware 3D architectures for GNN training is part of our future work.

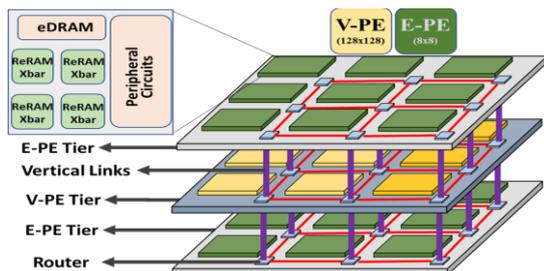

Fig. 2. ReGraphX architecture. This figure is for illustration purposes only.

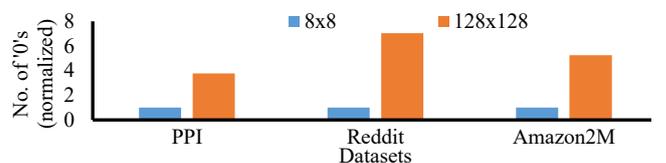

Fig. 3. Number of 'ZERO's stored (normalized with respect to that of 8x8) for two types of crossbars considering various datasets

### C. Pipelined GNNs on ReRAMs

Training GNNs on one big monolithic graph is often impractical due to memory concerns. To address this, clustering-based graph partitioning schemes are used. For instance, in [16], the authors use a graph partitioning tool (METIS [17]) to break a large graph into several smaller sub-graphs to reduce memory requirements. In addition, training on large graphs does not exploit the benefits of ReRAM-based architectures, which rely on a pipelined implementation [7]. Pipelining the different layers of a DNN reduces the number of ReRAM writes (which are slow), leading to higher overall throughput. However, this strategy is not amenable to GNNs if the entire graph needs to be processed altogether. Clustering/partitioning can be used to address this problem. By dividing the graph into clusters, we can implement the pipelined training strategy for GNNs (as shown in Fig. 4). Fig. 4 shows how pipelined training can be implemented for a GNN with two neural layers on ReRAMs as an example. As discussed earlier, each neural layer can be further divided into two sub-layers (E- and V-layers). Moreover, each E- and V-layer has a corresponding backward phase computation layer (e.g., BV1 represents the backward phase computation of V1 and so on). Here, V1 is the V-layer computation of the first neural layer of the GNN. Overall, this results in an eight-stage training pipeline as shown in Fig. 4.

The GNN pipelined training works as follows: First, we partition the large input graph into smaller sub-graphs. The size of each sub-graph is chosen based on training time, hardware requirements, and end-to-end accuracy. Here, each input-sub-graph (obtained after partitioning) is the equivalent of one input image in DNNs. The pipelined training strategy for GNNs can then be implemented analogous to DNNs: At time $T$ (Fig. 4), the first sub-graph ($G_1$) is loaded for V1-layer computations. The value of $T$ (in Fig. 4) will depend on the maximum of computation/communication times for any given layer. At the next timestamp $2T$, $G_1$ advances for subsequent E-layer computation ($E(G_1)$) while the next sub-graph $G_2$ is loaded for V1-layer computation, and so on. Once the pipeline is filled (at time $8T$), all the PEs (corresponding to all forward and backward phases) are active all the time, leading to higher throughput and hardware utilization [6], as shown in Fig. 4.

However, pipelined training also necessitates the simultaneous processing of multiple sub-graphs. For instance, at time $8T$ in Fig. 4, eight sub-graphs $G_1$-$G_8$ are being processed. As mentioned earlier in Sec. 3 (and Fig. 1(d)), processing each graph results in a many-to-one-to-many traffic pattern. Hence, processing multiple sub-graphs at the same time, as shown in Fig. 4, results in several sets of many-to-one-to-many traffic patterns corresponding to each of the sub-graphs $G_1$-$G_8$, resulting in high volume of data exchange which needs to be handled by the NoC. It is well known that in a pipeline, the slowest stage is the bottleneck and influences the overall execution time.

### D. Overall mapping to PEs

For an efficient implementation of pipelined GNN training, all the neural layers need to be executed simultaneously (Fig. 4). This requires keeping all the neural weights on the chip. Hence, we also need to allocate (map) adequate resources (ReRAMs) to each neural layer based on the requirements. In addition, the mapping strategy influences on-chip traffic and it must complement the underlying NoC design to ensure the best performance. In this work, we employ a simple Simulated Annealing (SA)-based strategy for optimal mapping following [12] as it can uncover high-quality solutions in a reasonable amount of time. The mapping of weights and the *Adj* matrix to the PEs can be envisioned as a combinatorial optimization problem: Given a total of $P$ PEs and $L$ layers (V and E), our aim is to distribute all the computation layers such that the highly communicating layers are mapped to nearby PEs. More specifically, our aim is to reduce long-range traffic (as much as possible), while ensuring efficient multicast communication.

## V. EXPERIMENTAL RESULTS

In this section, we first present the experimental setup to evaluate the performance of ReGraphX. Here, we demonstrate the NoC and the full system performance evaluation of the ReGraphX architecture.

### A. Experimental Setup

ReGraphX is a heterogeneous architecture that consists of two types of ReRAM tiles (V-PEs and the E-PEs, as shown in Fig. 2). The ReRAM crossbar and tile configurations (shown in Table I) of both the V- and E-PEs are based on [6] and [8] respectively. We consider 64 V-PEs (1 planar tier) and 128 E-PEs (distributed over 2 planar tiers). It should be noted that for both E-PEs and the V-PEs, the area of the ReRAM tiles is dominated by the peripheral circuits instead of the crossbars. The difference between E-PE and V-PE area is not significant, despite the difference in ReRAM crossbar structure. Moreover, for the sake of simplicity in physical floor planning, we assume that the E-PEs and V-PEs are spread over same planar footprint. The ReRAM tiles communicate with each other via the 3D NoC.

To evaluate the characteristics of the ReGraphX architecture, we use the performance models from [6].

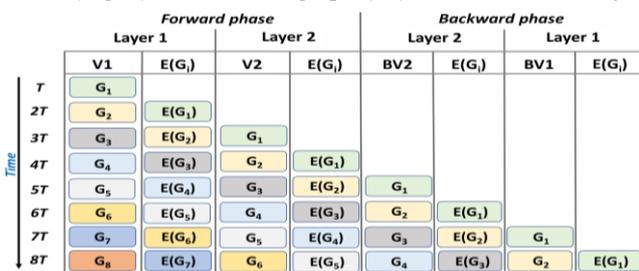

Fig. 4. Pipelined implementation of a 2-layer deep GNN, with forward and backward phase. Each layer has two-sublayers – V-layer and E-layer.

TABLE I. PARAMETERS OF THE REGRAPHX ARCHITECTURE [6] [8]

| V-PE: 1 Planar Tier, 64 Routers per tier, 4 Tiles per router | |
|---|---|
| ReRAM Tile (12 IMAs) | 1 IMA has: 8-ADCs (8-bits), 128x8 DACs (1-bit), 8 crossbars, 128x128 crossbar size, 10MHz, 2-bit resolution |
| E-PE: 2 Planar Tiers, 64 Routers per tier, 4 Tiles per router | |
| ReRAM Tiles (12 IMAs) | 1 IMA has: 8-ADCs (6-bits), 8x8 DACs (1-bit), 8 crossbars, 8x8 crossbar size, 10MHz, 2-bit resolution |

TABLE II. GRAPH DATA STATISTICS & GNN HYPER PARAMETERS

| Datasets | No. of Nodes | No. of Edges | No. of Partitions (NumPart) | Batch Size ($\beta$) | No. of Inputs (Num Input) |
|---|---|---|---|---|---|
| PPI | 56,944 | 818,716 | 250 | 5 | 50 |
| Reddit | 232,965 | 11,606,919 | 1500 | 10 | 150 |
| Amazon2M | 2,449,029 | 61,859,140 | 15000 | 10 | 1500 |

ReRAM arrays always execute instructions in-order and the instruction latencies are deterministic [6]. Hence, deterministic models have been used to evaluate ReRAM execution time, on-chip traffic, etc. [7]. The mapping of DNN layer weights and matrices on the tiles are determined offline. The traffic across the NoC is also statically determined to ensure conflict-free routing. We do not discuss the ReRAM execution models in detail for the sake of brevity and as it has been elaborated in [6].

In this work, we use the popular graph convolutional network (GCN) algorithm from [16] (implemented in TensorFlow), as the representative GNN for performance evaluation. However, our findings and the proposed architecture are equally applicable to other GNNs that rely on the recursive neighborhood aggregation scheme. The GCN configuration in our experiments uses graph partitioning to reduce memory overhead and enable pipelined training. This allows us to evaluate large-sized graphs that are otherwise impossible to process with limited memory, specifically in an on-chip environment. For the evaluation of GNN training on the ReGraphX architecture, we choose three popular graph datasets – *PPI, Reddit,* and *Amazon2M* (details provided in Table II). The GCN configuration for each dataset consists of four neural layers. As Table II shows, the datasets are diverse in nature (in terms of size, partitioning, etc.), which allows us to comprehensively evaluate the performance of ReGraphX.

### B. Effect of GNN hyper-parameters on ReGraphX

In this sub-section, we explore how various GNN hyper-parameters (particularly batch size) affect ReGraphX performance. Fig. 5 shows the training (Fig. 5(a)) and validation accuracy (Fig. 5(b)) when different batch size ($\beta$) is used for the *Reddit* dataset as an example. Note that the notion of $\beta$ in GNNs is not the same as in traditional DNNs. In a GNN, partitioning a graph (as in [16]) yields several smaller sub-graphs (referred as *NumPart* hereafter). However, this often leads to the loss of important information (graph connections) that can cause the training to be unstable. Hence, a stochastic multi-clustering approach is typically used, which involves merging back some of the *NumPart* sub-graphs. The number of sub-graphs that are merged back together to create an intermediate input sub-graph is defined as Batch Size ($\beta$) in GNNs. Hence, the number of effective input sub-graphs (*NumInput*) to the GNN is obtained by dividing *NumParts* by $\beta$. In Fig. 5, we chose *NumPart* as 1500 (following [16]) while varying $\beta$ from 1 to 20. From Fig. 5, it is clear that the choice of batch size does not affect the accuracy of the GNN significantly for the *Reddit* dataset. Similar observations were made for the other two datasets. However, it should be noted that smaller $\beta$ affects convergence, leading to unstable GNN training. For instance, as shown in Fig. 5, $\beta$ =1 and $\beta$ = 5 result in sudden drops in accuracy, even after a sufficiently high number of training epochs. On the other hand, larger $\beta$ leads to more stable training and should be preferred.

However, larger $\beta$ is costly to implement from the perspective of hardware overhead. Fig. 6 shows effects of $\beta$ on both the training time and hardware requirements (more specifically, E-PE requirements) in ReGraphX. A smaller value of $\beta$ results in smaller input-subgraphs, which in turn necessitates fewer E-PEs for storage. However, it leads to higher *NumInput* (i.e., more input sub-graphs) that need to be processed one-after-another. As shown in Fig. 6, this leads to higher training time. On the other hand, with increase in $\beta$, *NumInput* reduces, which in turn causes training time to decrease. However, higher $\beta$ leads to a drastic increase in E-PE requirements as larger graphs need to be stored on-chip. Interestingly, we note from Fig. 6 that the reduction in training time is relatively insignificant beyond $\beta$ = 10 while E-PE requirements keeps increasing steadily. From both Fig. 5 and Fig. 6, we note that larger $\beta$ leads to faster and more stable training, which is desirable. However, it also necessitates more E-PEs. Hence, we choose the maximum possible $\beta$ whose E-PE requirements can be met by ReGraphX specifications (Table I). In this case of *Reddit,* we set $\beta$ = 10. The values of $\beta$ for the other datasets are chosen following same methodology and are listed in Table II.

### C. NoC Evaluation

Next, we evaluate the performance of the 3D NoC, which serves as the communication backbone for ReGraphX. As described earlier (Fig. 4), training on ReRAMs is implemented in a pipelined fashion. Hence, the overall execution time will be dominated by the slowest computation/communication stages among all the layers. Fig. 7 shows the worst-case computation and communication delay when a GNN is trained on ReGraphX. The computation delay is the maximum time necessary to finish all the MAC operations of a given layer. Similarly, the communication delay shown in Fig. 7 represents the maximum time needed to finish sending all output data (messages) between the communicating neural layers.

As mentioned in Sec. IV, GNN exhibits many-to-one-to-many and multicast nature of traffic. Fig. 7 compares the performance when both unicast (denoted as Communication-U) and multicast routing (Communication-M) are used in ReGraphX . It also compares the communication delay with the computation delay on ReRAMs. Fig. 7 shows that for all datasets, the communication delay always dominates the overall pipeline stage latency, irrespective of the multicast support. As a result, the overall pipeline stage latency will be dictated by the relatively slower communication. Any further speed-up in computation will be meaningless as communication will not be able to keep up with it. This happens as the massive amount of multicast and many-to-one-to-many traffic between the PEs overwhelms the system, creating a performance bottleneck. Without 3D multicast support, the communication delay is 57.3% worse on average (Communication-U in Fig. 7). The communication delay improves significantly when multicast support is incorporated. It should be noted that for the Amazon dataset, the gap between computation and communication is almost non-existent. This happens as the *Amazon2M* dataset has a higher number of nodes, but smaller weight matrices compared to both *PPI* and *Reddit.* This results in higher computation delay, but the smaller weight matrices result in a relatively lower number of messages. Hence, the difference between communication and computation delay reduces. Multicast is used in the ReGraphX architecture hereafter because it helps in reducing the overall pipeline stage delay. This in turn results in lower overall execution time and higher speed-up compared to conventional GPU-based training.

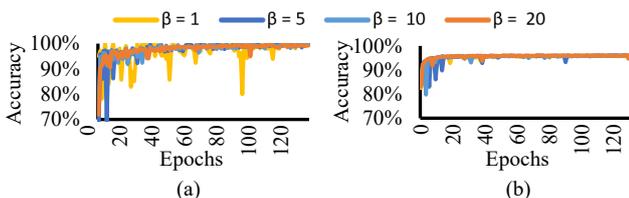

Fig. 5. GNN accuracy for different batch sizes for the Reddit dataset: (a) training; (b) validation.

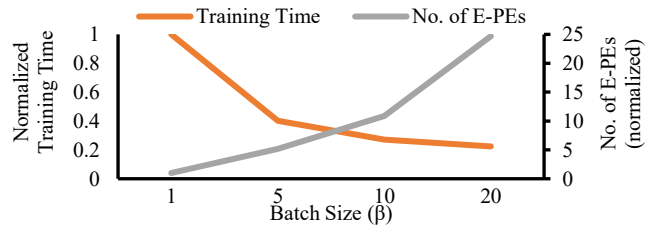

Fig. 6. Normalized training time and NumInputs for different Batch Size ($\beta$) for the Reddit dataset. All numbers have been normalized w.r.t $\beta$=1.

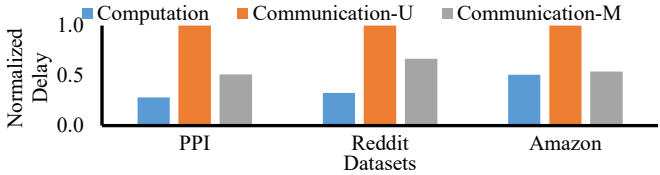

Fig. 7. Computation and Communication delay for ReGraphX

*D. Full System Evaluation*

Next, we undertake a full system performance evaluation and compare the overall execution time and energy dissipation of ReGraphX and a conventional GPU-based platform. For the GPU execution, we implement the Cluster-GCN algorithm [16] on NVIDIA Tesla V100 GPU. Fig. 8(a)-(c) show the normalized execution time, energy, and energy-delay-product (EDP) for GNN training on the ReGraphX architecture and the GPU, respectively. On an average, ReGraphX achieves 3X lower overall execution time on an average compared to the GPU. This speed-up can be attributed to: (a) ReRAM-based high-throughput MAC operations in ReGraphX, and (b) the 3D multicast-enabled NoC that reduces the communication latency. Fig. 8(b) shows that the ReGraphX architecture is on an average 11X more energy efficient than conventional GPUs due to the lower energy consumption of the ReRAM tiles. Overall, ReGraphX improves the EDP by 34X on an average and up to 40X compared to the conventional GPU implementation.

VI. CONCLUSION

Graph neural networks (GNNs) are being increasingly used in multitude of applications such as social media, drug discovery, and recommendation systems. In this work, we propose a novel heterogeneous ReRAM based architecture –

ReGraphX, tailor-made to accelerate GNN training. The ReRAM based computation platform is complemented with an efficient 3D NoC. By incorporating efficient many-to-one-to-many and multicast communication, the 3D NoC in ReGraphX reduces the communication bottleneck inherent in GNN training. Overall, ReGraphX outperforms conventional GPUs by up to 40X in terms of achievable EDP. This happens due to faster and more energy-efficient MAC computation enabled by the ReRAM crossbars and efficient communication enabled by the 3D NoC.

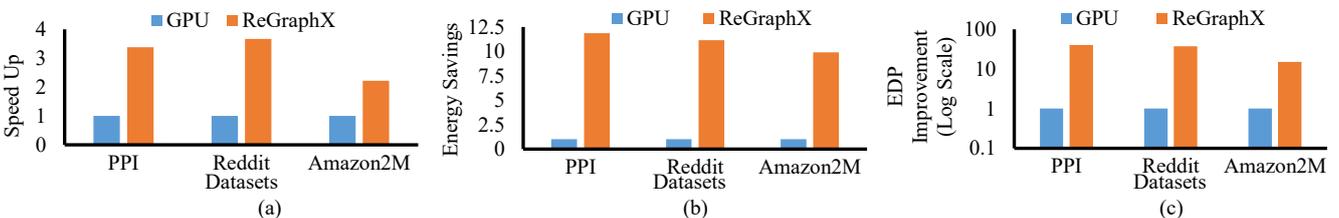

Fig. 8. ReGraphX (a) Execution speed-up, (b) Energy savings and (c) EDP Improvement compared to GPU (normalized with respect to GPU)